\documentclass[12pt]{article}
\usepackage[margin=1in]{geometry}
\usepackage{setspace}
\usepackage{lineno}
\usepackage{graphicx}
\usepackage{amsmath, amssymb}

\usepackage{graphicx,color}
\usepackage{epstopdf}
\usepackage{verbatim}
\usepackage{siunitx}

\usepackage{gensymb}
\usepackage{soul}

\usepackage{amsmath}
\usepackage{mathrsfs}
\usepackage{enumitem}
\usepackage{bm}
\usepackage{caption}
\usepackage{subcaption}
\usepackage{mathtools}
\usepackage{gensymb}
\usepackage{hyperref}
\usepackage[percent]{overpic}
\usepackage[overload]{empheq}

\usepackage{mathtools, cuted}

\usepackage[superscript]{cite}

\title{\bf Grid congestion stymies climate benefit from\\ U.S.\ vehicle electrification}

\author{
\hspace{-6mm} Chao Duan$^{1}$ and
Adilson E. Motter$^{2,3,4,5,*}$\\
\\
\hspace{-6mm}\normalsize{$^{1}$School of Electrical Engineering, Xi'an Jiaotong University, Xi'an 710049, China}\\
\hspace{-6mm}\normalsize{$^{2}$Department of Physics and Astronomy, Northwestern University, Evanston, IL 60208, USA}\\
\hspace{-6mm}\normalsize{$^{3}$Center for Network Dynamics, Northwestern University, Evanston, IL 60208, USA}\\
\hspace{-6mm}\normalsize{$^{4}$Department of Engineering Sciences and Applied Mathematics,} \\ \normalsize{Northwestern University, Evanston, IL 60208, USA}\\
\hspace{-6mm}\normalsize{$^{5}$Northwestern Institute on Complex Systems, Northwestern University, Evanston, IL 60208, USA}\\
\medskip
\hspace{-6mm}\normalsize{$^{*}$ Corresponding author:  motter@northwestern.edu}
}

\date{}

\begin{document}

\maketitle

\begin{abstract}
\noindent\bf
Averting catastrophic global warming requires decisive action to decarbonize key sectors. Vehicle electrification, alongside renewable energy integration, is a long-term strategy toward zero carbon emissions. However, transitioning to fully renewable electricity may take decades---during which electric vehicles may still rely on carbon-intensive electricity. We analyze the critical role of the transmission network in enabling or constraining emissions reduction from U.S. vehicle electrification. Our models reveal that the available transmission capacity severely limits potential CO$_2$ emissions reduction. With adequate transmission, full electrification could nearly eliminate vehicle operational CO$_2$ emissions once renewable generation reaches the existing nonrenewable capacity. In contrast, the current grid would support only a fraction of that benefit. Achieving the full emissions reduction potential of vehicle electrification during this transition will require a moderate but targeted increase in transmission capacity. Our findings underscore the pressing need to enhance transmission infrastructure to unlock the climate benefits of large-scale electrification and renewable integration.
\end{abstract}
\medskip

\section*{Introduction}

The 2015 Paris Agreement, a landmark in multilateral climate policy, set the goal to limit global warming to well below 2 \degree C (preferably below 1.5 \degree C) compared to pre-industrial levels \cite{streck2016paris}. To accomplish this goal, nations must take bold actions to curb carbon emissions as soon as possible and achieve carbon neutrality by mid-century. Studies on the existing climate policies show that the time window for achieving the 1.5 \degree C goal may have closed \cite{rogelj2015energy,rogelj2016paris} and the world is heading for 3 \degree C of warming by the year 2100  \cite{rogelj2016paris,hausfather2020emissions}. Thus, substantial upgrading of current policies is required to reverse the trend toward catastrophic outcomes. Electricity plays a central role in the decarbonization of the energy and other infrastructure industries \cite{williams2012tech,chu2012opportunities}. A widely recognized pathway to zero emissions includes the transition to electricity generation from renewable sources combined with the electrification of energy-intensive sectors currently powered by fossil fuels, including transportation, heating, and cooling \cite{sachs2016pathways,davis2018net}. However, due to techno-economic barriers, the ongoing transition to renewable energy generation will likely take decades to complete, and during this period, electricity must be generated from a mix of fossil and renewable sources \cite{bogdanov2019radical,luderer2021impact}. At different stages of this transition, the carbon reduction achieved by electrification can vary widely. It has been shown that electrification of urban infrastructures becomes carbon competitive when the grid-average emissions intensity is lower than about $600$ tCO$_2$ per GWh (tonnes of CO$_2$ emitted per gigawatt-hour of electricity generated, where weights are in the metric system throughout)\cite{Kennedy2013low}. Recent studies indicate that the overwhelming majority of the world's transportation and heating demand can already derive climate benefits from electrification under the current generation mixes \cite{knobloch2020net}, and that retiring a small fraction of most polluting generation units worldwide could lead to substantial additional climate benefits from electrification \cite{tong2018target}. Therefore, there is no question today about the benefits of electrifying the energy-intensive sectors. The outstanding question is how to maximize the benefits of electrification amidst the decades-long transition from fossil fuels to renewable energy, which we address here by focusing on the electrification of the transportation sector in the United States.
 
The transportation sector is responsible for 29\% of U.S. greenhouse gas emissions,  81\% of which come from light-duty and heavy-duty motor vehicles \cite{hockstad2018inventory}. 
Research shows that, in the U.S., current commercial electric vehicle (EV) technology can meet the energy requirements 
of the bulk
of vehicle-days without recharging, supporting the feasibility of widespread electrification of passenger vehicles \cite{needell2016potential}. The extent to which vehicle electrification contributes to carbon reduction depends on the CO$_2$ emissions from the energy generation technologies charging the EVs. The average fuel efficiency of conventional cars made in 2017 is 26.0 MPGe (miles per gallon equivalent), whereas the efficiency of EVs from the same year is 98.2 MPGe. 
Thus, a gallon
of gasoline consumed by conventional cars, which generates 8.9 kg of CO$_2$,
amounts to an average of 8.9 kWh of electricity demand from EVs.
This implies that, for each gallon of gasoline consumed by a conventional car, switching to an EV would bring a reduction of roughly $8.9\,(1-x\cdot 10^{-3})\,$kg of CO$_2$ emissions if the EV is charged in a power network with an emissions intensity of $x$ tCO$_2$ per GWh.
This work examines the electrification of
gasoline-powered vehicles, which consist mainly of light-duty vehicles but also include medium- and heavy-duty ones. 
The energy efficiency ratio between  electric and conventional heavy-duty vehicles and the corresponding reduction in emissions are
comparable and assumed to be the same as for light-duty vehicles \cite{california2018battery}---e.g., in the  U.S.,
electric trucks
currently emit at least $42\%$--$61\%$ less CO$_2$ from electricity generation than diesel trucks emit from fuel combustion \cite{lee2013electric}. 
In highly populated regions, such as some areas of China and Europe, vehicle electrification can also deliver substantial air quality and health benefits independently of the net climate benefits \cite{liang2019air,he2022}. Thus, as the grid penetration of solar, wind, and other renewable sources increases, the climate and health benefits of vehicle electrification are projected to grow substantially, meaning less carbon emissions and fewer premature deaths worldwide \cite{wu2012energy,tessum2014life,peters2020public}.

Another dimension of the benefits of the transition to EVs comes from the interplay between EV charging and power grid operation. Due to the intermittency of solar and wind energy generation, a major factor that determines how much renewable energy can be integrated into the system is the amount of generation and demand flexibility available in the grid. 
Large-scale vehicle electrification can provide such flexibility by 
allowing the modulation of EV battery charging and discharging to help mitigate the intermittency associated with renewable energy generation \cite{richardson2013electric,coignard2018clean,bellocchi2019role,zhang2020quantifying}. Therefore, EVs enable increased integration of clean energy, which helps decrease the grid-average emissions intensity and amplifies carbon reduction achieved by vehicle electrification. In practice, the realization of the above-mentioned benefits of vehicle electrification could depend on many factors that have been analyzed in the literature, including marginal generation mix \cite{mccarthy2010determining}, scale and types of EVs \cite{thomas2012green,requia2018clean}, availability of charging infrastructure \cite{requia2018clean}, unidirectional vs.\ bidirectional charging \cite{coignard2018clean,zhang2020quantifying}, and incentives promoting controlled charging \cite{sierzchula2014influence}. 
However, surprisingly scarce attention has been given to a more fundamental factor in determining the climate benefits of vehicle electrification, namely the power transmission network \cite{brummitt2013,witthaut2022}. 

We posit that without a physically adequate grid infrastructure, the portion of available renewable energy that can be delivered to charge EVs will be limited by the transmission capacity of the system. Recent reports from the U.S. Department of Energy \cite{DOEvehicles,DOEimpact} underscore that widespread EV adoption presents both challenges and opportunities across all levels of the power system---distribution, subtransmission, and transmission---highlighting the need for comprehensive analysis at both local and system-wide scales. In addition, a report from the National Center for Energy Analytics \cite{lesser2024infrastructure} predicts that electric utilities will have to upgrade their distribution
systems to accommodate the increased demand from EV charging. 
Although these authoritative reports identify critical research gaps and a general need for infrastructure upgrades, they offer limited quantitative insights and explicitly call for further investigation.
To help advance this investigation, we present a quantitative analysis of how transmission constraints may limit the CO$_2$ reductions enabled by vehicle electrification, while also identifying infrastructure upgrades to maximize these reductions.
This is achieved by devising a full-scale power transmission network and implementing three power flow optimization models: Model I (pre-electrification baseline), Model II (vehicle electrification impact), and Model III (optimal grid upgrades for CO$_2$ reduction).
Our study focuses on vehicle electrification and power grid operation across the contiguous
United States.

\section*{Results}

\subsection*{The U.S.\ power system.}
\begin{figure*}[t]
\begin{center}
      \begin{overpic}[width=1.0\textwidth,tics=5]{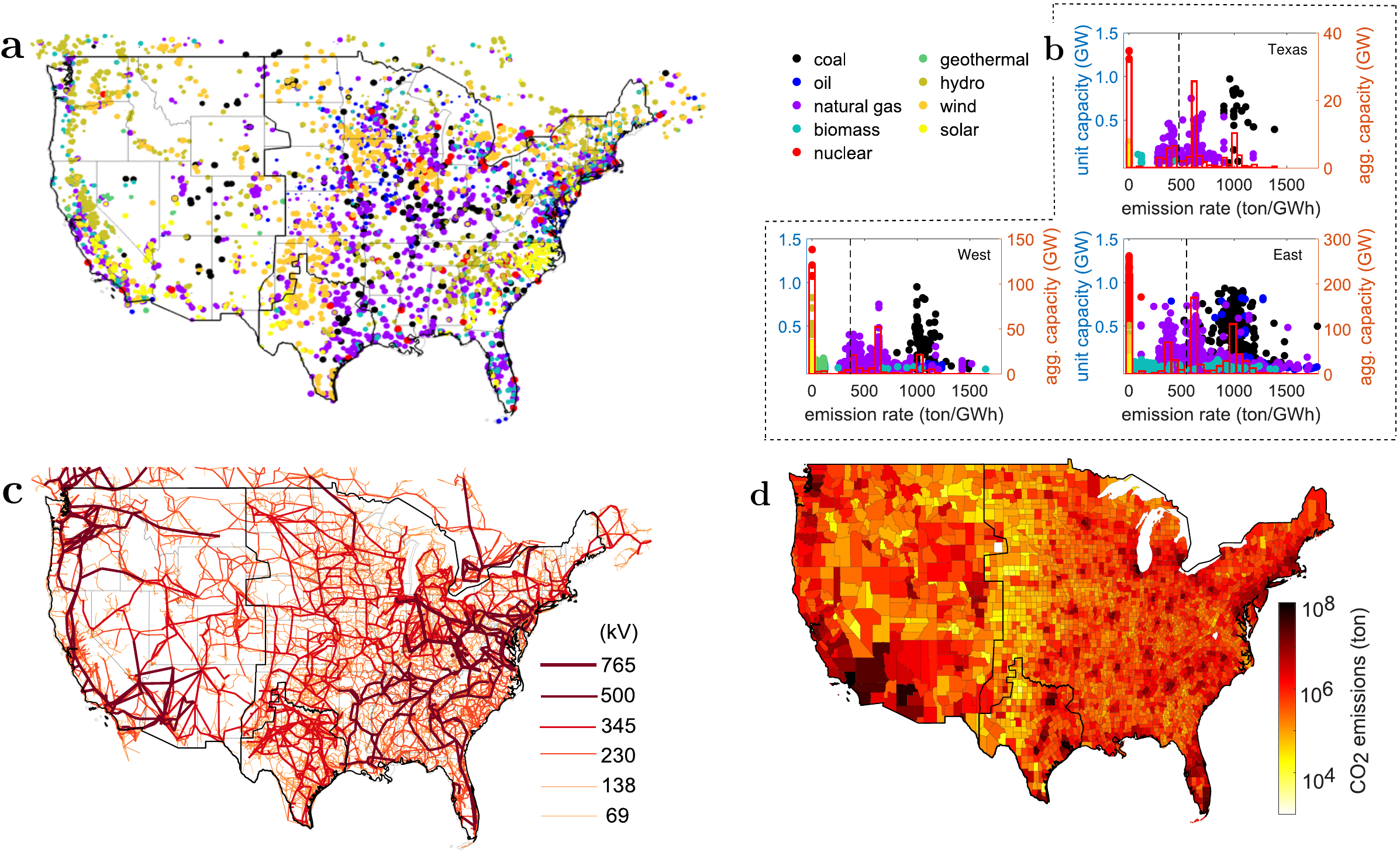}
\end{overpic}
\caption{\textbf{Overview of the U.S. generation sources, power grid, and vehicle emissions in 2018.}  \textbf{a},~Geographical distribution of generation units color-coded according to the types of primary energy source across all three interconnections of the U.S. power system (including extensions into southern Canada). \textbf{b},~Scatter plots  (left axes) and histograms   (right axes) of the capacity of the generation units versus emission rates, where the color code is the same as in panel (a). The dashed line marks the average emission rate, which is $364$, $475$, and $547$ t per GWh for the Western, Texas, and Eastern interconnections, respectively. \textbf{c},~Contiguous U.S. power grid, represented as a complex multilayer network of power lines operating at different voltage levels.
\textbf{d},~County-level distribution of $\text{CO}_2$  emissions from motor vehicles, estimated from the state-wise fuel consumption reported in the Highway Statistics \cite{FHA}. This estimation follows from distributing state-level emissions in proportion to the population of each county.}
\label{uspowergrid}
\end{center}
\end{figure*}

The contiguous U.S. power system consists of three major networks---the Western, Texas, and Eastern interconnections, each of which contains a diverse generation mix of fossil-fuel and renewable sources (Fig.~\ref{uspowergrid}a).
Like other energy systems in the world, the U.S. system is undergoing a transition from carbon-intensive to renewable energy. However, the three interconnections have been at substantially different stages of renewable energy integration. In 2018, the installed capacity for the major renewable generation, comprising solar, wind, and hydropower, corresponded to $37.3\%$, $23.2\%$, and $13.3\%$ of the total generation capacity in the Western, Texas, and Eastern interconnections, respectively. The different levels of renewable energy penetration lead to different distributions of CO$_2$ emission rates (Fig.~\ref{uspowergrid}b). The 2018 average CO$_2$ emission rates in the three interconnections were respectively $364$, $475$, and $547$ t per GWh, which indicates that they all have passed the threshold for carbon competitive electrification of fossil-fuel powered sectors \cite{Kennedy2013low}. Therefore, the transition from internal combustion vehicles (ICVs) to EVs would certainly result in carbon reduction for the transportation sector. However, the quantitative analysis of how much CO$_2$ can be reduced requires consideration of the geographical and electrical distributions of the generation and demand, as well as the spatial and temporal variability of renewable sources. The additional electricity demand arising from vehicle electrification will likely be drawn from the power grid nodes near where the energy is finally consumed by the EVs (Fig.~\ref{uspowergrid}c). According to Highway Statistics for the year 2018, the annual motor fuel consumption in the U.S. was about 194 billion gallons of gasoline or equivalent, which corresponds to about 1.72 billion tons of CO$_2$ emissions. The motor fuel consumption and carbon emissions are unevenly distributed across the country (Fig.~\ref{uspowergrid}d). With the motor fuel replaced by electric power drawn locally from the grid, the incurred carbon emissions are determined by the complex interaction among the sources (Figs.~\ref{uspowergrid}a-b), the transmission network (Fig.~\ref{uspowergrid}c), and the demand pattern (Fig.~\ref{uspowergrid}d).

\subsection*{Energy flow modeling.}
To model the effects of grid constraints on vehicle electrification, we need to consider 1) the physical structure of the power system, 2) the spatial and temporal variability of the renewable generation and power demand \cite{zeyringer2018designing,macdonald2016future}, and 3) the interaction and cooperation between EV charging and the dispatch of power generation \cite{williams2012tech,zhang2020quantifying}. Based on the U.S. power grid data from the Federal Energy Regulatory Commission (FERC), we built three power flow optimization models considering all aspects mentioned above. The three models respectively describe the power system operation before vehicle electrification (Model I), the change of system state induced by a given level of vehicle electrification (Model II), and an optimal strategy for transmission-line upgrading that reduces CO$_2$ emissions (Model III). In addition to the FERC power grid data, the models incorporate regional hourly power demand data from the U.S. Energy Information Administration (EIA), location-wise hourly wind and solar generation data from the National Renewable Energy Laboratory (NREL), power plant emission rate data from the Environmental Protection Agency (EPA), and state-wise motor vehicle fuel consumption data from Federal Highway Administration (FHWA). 

To account for vehicle electrification, we assume that the fuel efficiency ratio between ICVs and EVs is $26.0$:$98.2$, which is consistent with average real-world efficiencies given in the EPA 2019 Automotive Trends Report \cite{EPAtrend}.  Under this assumption, for the same travel distance, one gallon of gasoline consumed by an ICV
amounts to $8.9$ kWh of electricity demand from an EV. We distribute the motor fuel consumption of each state to its counties in proportion to the county population (Fig. \ref{uspowergrid}d), and we assume that the EV charging stations of each county are connected to power grid nodes located in that county with voltage levels lower than $200$~kV. This way, for any given percentage of vehicle electrification, we can estimate the daily electricity demand from EVs in each county. Model II determines the optimal charging strategies that minimize carbon emissions from power plants after the integration of EVs (optimal charging scheduling is also desirable to avoid the risk of voltage violations \cite{Helou2022}). By comparing the results from Models I and II, we can calculate the net effects of vehicle electrification. For full details of the mathematical models and data compilation, see Methods.

\subsection*{Congestion-induced CO$_2$ emissions.}
\begin{figure}[t]
\begin{center}
\begin{overpic}[width=0.85\textwidth,tics=5]{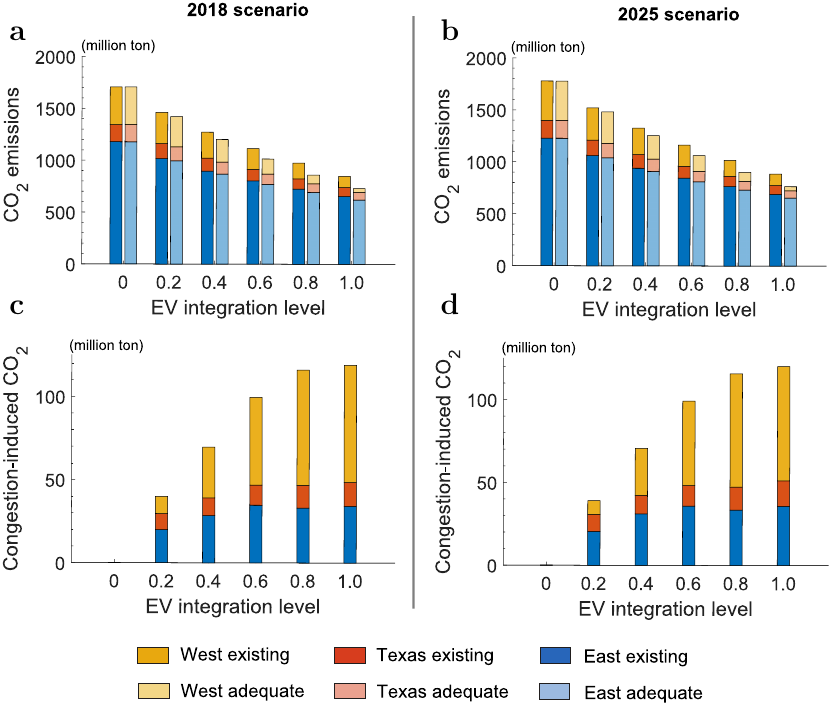}
\end{overpic}
\caption{\textbf{ Vehicle operational CO$_2$ emissions vs.\ EV integration.}  \textbf{a, b},~Annual vehicle operational $\text{CO}_2$ emissions from motor vehicles as a function of EV integration levels for the 2018 scenario (\textbf{a}) and 2025 scenario (\textbf{b}). These emissions include tailpipe emissions from ICVs and electricity generation emissions to power EVs; the integration level is the fraction of vehicles replaced by EVs.
The three stacked segments of each bar represent the portions contributed by the Western (orange),  Texas (red), and  Eastern (blue) interconnections, respectively. A comparison is drawn between the realistic scenario given by the existing transmission capacity (darker color bars) and an ideal scenario with adequate transmission capacity (lighter color bars). \textbf{c, d},~Annual $\text{CO}_2$ emissions induced by network congestion as a function of the EV integration levels for the 2018 scenario (\textbf{c}) and 2025 scenario (\textbf{d}). The 2025 scenario is modeled according to the generation mix projection from EIA.
}
\label{fig:carbonreduce2020}
\end{center}
\end{figure}

We consider different levels of vehicle electrification from $0\%$ to $100\%$ and solve Models I and II to obtain the amount of {\it vehicle operational CO$_2$ emissions}, which comprises the tailpipe emissions from ICVs and the grid emissions from power plants generating the electricity used by EVs. Our analysis focuses solely on operational emissions, excluding those from manufacturing, fuel production, and other life-cycle stages.
Under the generation mix of 2018, we observe a consistent reduction of vehicle operational CO$_2$ emissions in all three interconnections as the percentage of EVs increases (Fig.~\ref{fig:carbonreduce2020}a). When all ICVs are replaced by EVs, the vehicle operational CO$_2$ emissions in the U.S. drop from $1707$ Mt per year (million tons per year) to $849$ Mt per year---an overall reduction of $50.3\%$. The three networks contribute very differently to the national carbon reduction due to disparate levels of integration of renewable energy. Complete vehicle electrification leads to $70.0\%$, $49.1\%$, and $41.3\%$ reduction in vehicle operational carbon emissions in the Western, Texas, and Eastern systems, respectively. However, further reduction of carbon emissions is limited not only by the generation mix but also by the delivery capacity of the network. A network is said to have adequate capacity when the operation state that minimizes CO$_2$ emissions is not limited by the capacity of the transmission lines (and thus the network is uncongested) at any point in time.
With adequate network capacity, the Western, Texas, and Eastern systems would have achieved $89.0\%$, $58.4\%$, and $44.4\%$ reduction in vehicle operational CO$_2$ emissions with full vehicle electrification under the 2018 generation mix (Fig.~\ref{fig:carbonreduce2020}c).
The effect of network capacity constraints is most prominent in the Western interconnection because this network has the highest penetration level of renewable energy.
In other words, as motor vehicles are electrified, part of the CO$_2$ emissions is purely caused by network congestion, which we refer to as congestion-induced CO$_2$ emissions. As shown in Fig.~\ref{fig:carbonreduce2020}c, the congestion-induced CO$_2$ emissions increase with the EV integration levels and finally reach $118.9$ Mt per year, corresponding to an emission overhead of $16.3\%$. Similar trends are observed under the projected 2025 generation mix and number of vehicles (Figs.~\ref{fig:carbonreduce2020}b and \ref{fig:carbonreduce2020}d). In this case, the congestion-induced CO$_2$ emissions with full EV integration are estimated to be $120.3$ Mt per year (an overhead of $15.7\%$), which is only modestly lower in percentage than in 2018 despite the projected increase in renewable energy penetration.

The reported scenarios serve as diagnostic tools rather than forecasts, helping to isolate infrastructure effects and provide policy-ready benchmarks.
The use of 2018 grid data may introduce discrepancies between the studied and real scenarios, as the grid has undergone upgrades since then.
However, the rate of construction of new high-voltage transmission lines in the U.S.\ has been slow over this period, with an average of only 680 miles added annually between 2015 and 2023 \cite{FEWer}---a modest change relative to the more than 500,000 miles of existing high-voltage transmission lines.

\begin{figure*}[t]
\begin{center}
\begin{overpic}[width=0.85\textwidth,tics=5]{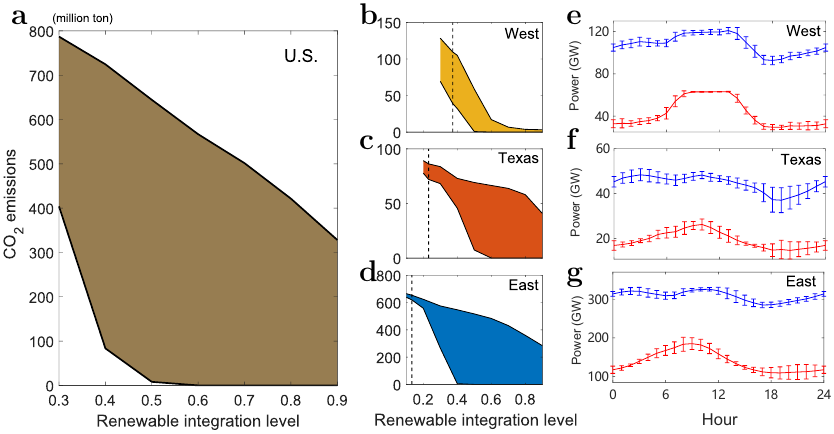}
\end{overpic}
\caption{\textbf{Annual CO$_2$ emissions vs.\ renewable integration.} \textbf{a}, Vehicle operational CO$_2$ emissions plotted against the renewable integration level when assuming $100\%$ vehicle electrification in the entire U.S. grid.
The upper edge of the shaded area represents the CO$_2$ emissions with the actual 2018 network capacity constraints, whereas the lower edge represents the emissions in the 
congestion-free case. The renewable integration level is the fraction of generation capacity powered by solar, wind, and hydroelectric plants.
\textbf{b}-\textbf{d}, Breakdown of (a) into the Western (\textbf{b}), Texas (\textbf{c}), and Eastern (\textbf{d}) networks.  The dashed lines indicate the 2018 integration levels (corresponding to a full-system integration level of $20.1\%$). \textbf{e}-\textbf{g}, Average EV charging (red) and renewable generation (blue) for each interconnection over a 24-hour period, where the error bars represent the standard deviation across the average day of each month over a year. The scenario assumes 100\% EV penetration, no transmission constraints, and renewable integration levels of 50\% for the Western interconnection (\textbf{e}), 60\% for the Texas interconnection (\textbf{f}), and 40\% for the Eastern interconnection~(\textbf{g}).}

\label{fig:emission_vs_renew}
\end{center}
\end{figure*}

As the U.S. power system undergoes a transition toward higher penetration of renewable energy, it is clear that the level of penetration will strongly impact the carbon reduction achieved by vehicle electrification, but it is not clear how grid congestion will limit this impact.
Here, we continue our quantitative analysis by examining the vehicle operational and congestion-induced CO$_2$ emissions as functions of the level of renewable integration using the 2018 network as a reference. The integration level is varied for the  
three interconnections by scaling up and down the capacity of the existing solar and wind power plants while keeping the capacity of other sources fixed. We then solve our optimization Models I and II to predict how the vehicle operational CO$_2$ emissions are impacted by the change in the generation mix for a fixed $100\%$ of vehicle electrification. For increasing penetration of renewable energy, the vehicle operational CO$_2$ emissions reduce monotonically in 
the full system (Fig.~\ref{fig:emission_vs_renew}a) and also within
each regional network (Figs.~\ref{fig:emission_vs_renew}b-d). However, the carbon reduction enabled by renewable energy is severely limited by the available transmission capacity, rendering a large portion of the vehicle operational emissions purely congestion-induced. 

It follows from Figs.~\ref{fig:emission_vs_renew}b-d that, if it were not due to transmission limitations, the three interconnections could achieve zero vehicle operational CO$_2$ emissions with $40\%$-$60\%$ generation capacity from renewable energy. 
In other words, under $40\%$-$60\%$ renewable integration and no transmission constraints, the renewable energy that would otherwise be curtailed in a system without EVs is sufficient to fully power a $100\%$ electrified vehicle fleet.
Figures \ref{fig:emission_vs_renew}e-g show how renewable generation and optimal EV charging vary over time at these levels of renewable energy integration and EV penetration, illustrating their temporal correlation.
 
In particular, the full system would transition from $849$ Mt per year to less than $8.1$ Mt per year vehicle operational carbon emission when renewable energy penetration reaches $50\%$ (i.e., when it matches the existing non-renewable generation capacity) (Fig.~\ref{fig:emission_vs_renew}a). 
However, with the 2018 grid capacity, a renewable integration level of $50\%$ would still result in vehicle operational CO$_2$ emissions of $644$ Mt per year, over $98.7\%$ of which is caused by network congestion; this corresponds to a reduction of just $24.2\%$ relative to $849$ Mt per year (the 2018 scenario with full vehicle electrification)
and a reduction of $62.3\%$ relative to $1707$ Mt per year (the 2018 {\it status quo}).
Furthermore, with adequate transmission capacity, a renewable integration level of $30\%$ is already enough to reduce the vehicle operational CO$_2$ emissions to $400$ Mt per year. On the other hand, it would require over $80\%$ renewable integration to bring the CO$_2$ emissions down to the same level in the absence of a transmission upgrade.

\subsection*{Optimal grid upgrading.} 
\begin{figure}[t]
\begin{center}
\begin{overpic}[width=0.85\textwidth,tics=5]{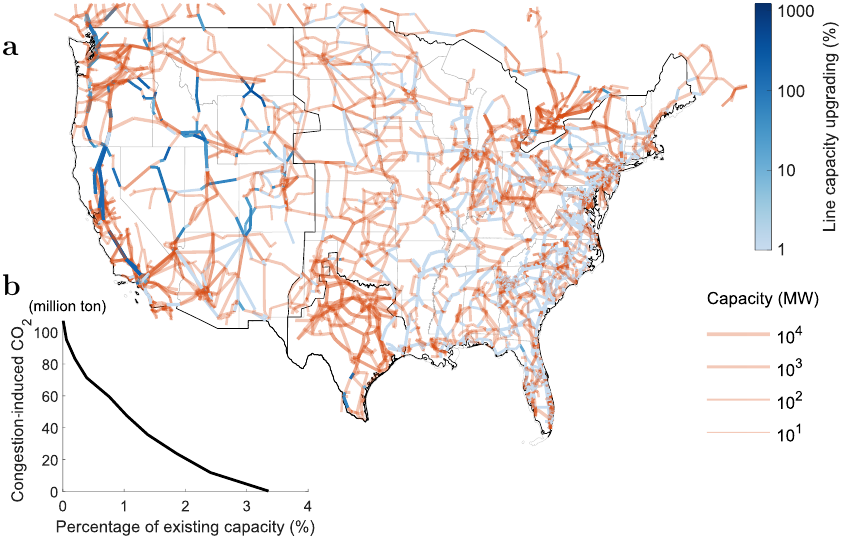}
\end{overpic}
\caption{\textbf{Optimal U.S. grid upgrading relative to the 2018 scenario.}  \textbf{a},~Minimal capacity upgrading of the U.S. transmission grid to eliminate congestion-induced $\text{CO}_2$ emissions assuming an EV integration of $100\%$.  Blue marks upgraded transmission lines, with the color gradient indicating the capacity increase as a percentage of the existing capacity. The other transmission lines (red) require no upgrading. The final capacity of each line is coded by the width of the line. \textbf{b},~Congestion-induced $\text{CO}_2$ emissions in the U.S. as a function of the transmission capacity upgrade measured as a percentage of the existing capacity. In both panels, we only consider the transmission lines with voltage levels above $200$ kV.}
\label{fig:upgrade2020}
\end{center}
\end{figure}

Given the analysis above, it is important to ask how much transmission line upgrading is needed to fully realize the climate benefit of vehicle electrification accompanied by renewable-energy integration. To address this question, we implement the optimization Model III (Methods) on the U.S. power grid with the 2018 generation mix and transmission capacity under the assumption of 100$\%$ vehicle electrification. 
This optimization model formulates transmission upgrade planning as a minimum-cost capacity allocation problem, where the decision variables are the incremental capacity upgrades $\mathrm{\Delta} F_l$ for each line $l$. The objective is to minimize the total network expansion costs while ensuring sufficient capacity to achieve the specified reduction in EV charging emissions. The solution must satisfy four key constraints: (i) real-time power balance; (ii) power flow limits incorporating capacity expansions; (iii) generator technical constraints, including capacity bounds and ramp rate limits; and (iv) guaranteed fulfillment of EV charging demand. The model simultaneously optimizes generation dispatch and charging schedules to meet these requirements. 
Since our focus is on the transmission network, we only consider the congestion and upgrading of transmission lines with voltage levels higher than $200$ kV. The results are presented in Fig.~\ref{fig:upgrade2020}. As more capacity is (optimally) added to existing transmission lines, the congestion-induced CO$_2$ emissions decrease monotonically to zero with diminishing marginal benefits (Fig.~\ref{fig:upgrade2020}b). The minimal line upgrade necessary to eliminate congestion is about $3.4\%$ nationwide. There are pronounced differences among the three interconnections. The necessary line capacity upgrading in the Western, Texas, and Eastern networks is respectively $7258$, $95$, and $55$ GW$\cdot$mile, which amounts to about $10.5\%$, $0.35\%$, and $0.04\%$ of the corresponding installed capacity. These differences are largely due to the disparate levels of renewable penetration across the three interconnections.

\begin{figure}[t]
\begin{center}
\begin{overpic}[width=0.85\textwidth,tics=5]{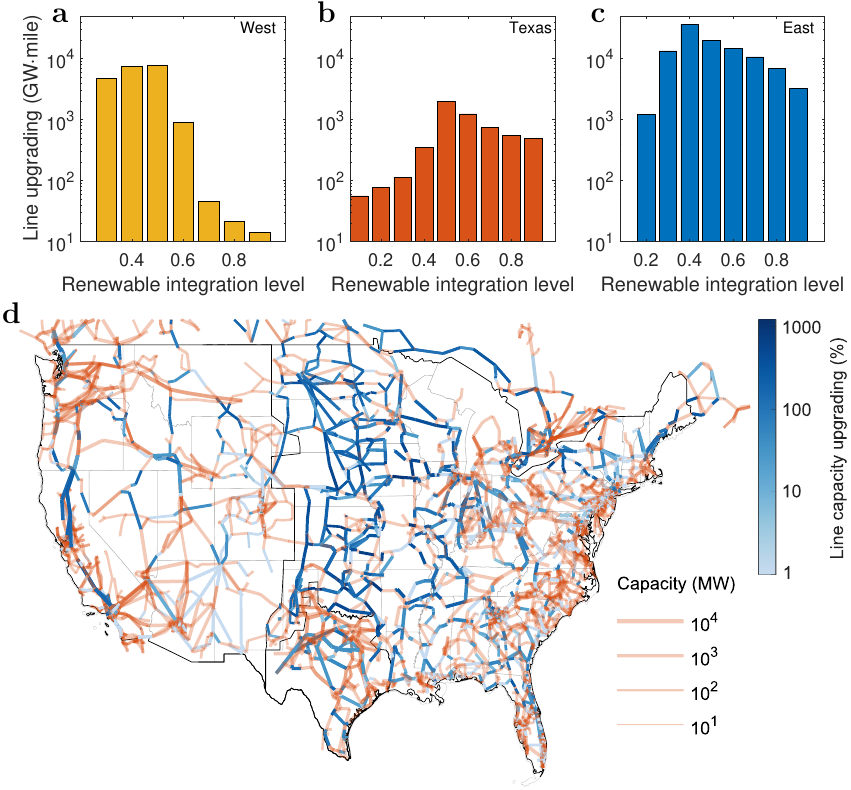}
\end{overpic}
\caption{\textbf{Optimal upgrading of the U.S.\  grid vs.\ renewable integration.}  \textbf{a}-\textbf{c},~Optimal amount of transmission line upgrading required to eliminate congestion-induced $\text{CO}_2$ emission as a function of the level of renewable integration in the Western (\textbf{a}), Texas (\textbf{b}), and Eastern (\textbf{c}) interconnections. 
The upgrades account for the maximum power flow on the corresponding line over the four seasons of the year while considering $100\%$ vehicle electrification.
The required upgrades peak at the renewable integration levels of approximately $50\%$ in the Western and Texas networks and $40\%$ in the Eastern network. 
\textbf{d},~Visualization of the peak of the required upgrade in
each interconnection using the same color and line schemes as in Fig.~\ref{fig:upgrade2020}a.
}
\label{fig:upgrade_max}
\end{center}
\end{figure}

The optimal line upgrades achieving zero congestion-induced CO$_2$ emissions for the 2018 generation mix are visualized in Fig.~\ref{fig:upgrade2020}a, showing that strategic upgrades
must be planned in a globally coordinated manner within each interconnection. 
The need for transmission line upgrading increases as more renewable generation is added to the grid. 
For the capacity constraints of 2018, the peak required upgrade is predicted to occur for a renewable integration level of approximately $50\%$ in the Western (Fig.~\ref{fig:upgrade_max}a), 
$50\%$ in the Texas (Fig.~\ref{fig:upgrade_max}b),
and $40\%$ in the Eastern
(Fig.~\ref{fig:upgrade_max}c)
interconnection, 
and thus they occur at nonidentical but comparable levels of renewable integration (cf.~Figs.~\ref{fig:emission_vs_renew}b-d).
The optimal capacity increases at these peaks are $7661$, $1966$, and $36778$~GW$\cdot$mile, corresponding to $11.1\%$, $7.1\%$, and $29.6\%$ of the respective 2018 existing capacities.
Nationwide, the peak of the required upgrade in each interconnection would lead to a capacity increase in $2490$ out of $13114$ transmission lines above $200$ kV, with the percentage increase very unevenly distributed among the upgraded lines (Fig.~\ref{fig:upgrade_max}d).
For a renewable integration level of $50\%$ in each interconnection, corresponding to a renewable generation capacity similar to the available non-renewable generation capacity, the overall line capacity upgrade required is predicted to be $13.4\%$.
A comparison between Fig.~\ref{fig:upgrade2020}a and Fig.~\ref{fig:upgrade_max}d reveals that while the required upgrades remain similar in some regions (e.g., California), they show significant variations in other regions (e.g., central states). These regional disparities stem from differing levels of renewable integration and infrastructure development across the three interconnections, highlighting the need for tailored upgrading strategies.

\section*{Discussion}

Our quantitative study reveals a significant limitation imposed by the existing U.S. grid infrastructure on the climate benefits that can be achieved from the ongoing vehicle electrification and renewable energy transitions. The analysis reinforces the pressing need to make a strategic effort to upgrade the transmission network
 far beyond current regional transmission planning associated with interconnection queues \cite{Armstrong2024}. Simply put, when the total renewable generation capacity reaches the existing non-renewable generation capacity, full vehicle electrification would essentially eliminate all the vehicle operational CO$_2$ emissions if the transmission grid had adequate capacity. However, for the transmission capacity available in the existing grid, full vehicle electrification under such high renewable energy penetration 
would result in significant congestion-induced CO$_2$ emissions.
Our analysis for the U.S.\ predicts that to eliminate congestion-induced emissions during the EV and renewable transitions, a transmission capacity increase (or construction) of specific power lines is required.
This demonstrates the importance of coordinated, system-wide planning, as increasing renewable capacity in one region often requires complementary transmission investments in other areas to realize full decarbonization benefits. Even though our analysis is based on data from the U.S., the conclusions likely extend to other countries, especially those that have long been under-investing in electricity transmission infrastructure.

The analysis in this study relies on a few assumptions regarding data, models, and human behavior. First, the spatiotemporal distributions of renewable generation and load demand for different renewable energy integration scenarios are derived by scaling up the base power flow values from FERC data according to the respective aggregate projections for each interconnection. Second, our power flow modeling accounts only for static transmission line capacity limits, omitting dynamic constraints such as synchronization, frequency, and voltage stability. Third, within each state, we assume that the vehicle fuel consumption and EV adoption rates are distributed in proportion to the population of each county. Future refinements could strengthen the conclusions---particularly by incorporating stability constraints in power flow modeling and accounting for uneven EV adoption. A more constrained network with spatially heterogeneous loads is expected to increase congestion-induced CO$_2$ emissions.
In contrast, it is less clear whether our assumptions about renewable generation and load distribution introduce any particular bias, even though higher-resolution data would likely yield a more accurate model.
In addition, we emphasize that our vehicle-specific upgrade targets represent lower-bound estimates. Future real-world transmission planning must integrate all forecast electrification loads (heating, cooling, industrial processes, data centers, etc.) using the framework we have established and combine these sectoral analyses into unified national infrastructure pathways.
In particular, the growth in energy demand associated with the proliferation of data centers for AI operations and cryptocurrency mining intensifies the planning challenges.  Distributed generation and strengthened power plant greenhouse gas standards \cite{EPAstandard} can partially mitigate the problem by alleviating grid congestion and reducing emissions, but they cannot replace the need for comprehensive power transmission modernization.

There is now consensus that electrification of transportation and other energy-intensive sectors, along with the increased integration of renewable energy, is a key path toward decarbonization.
However, this requires an adequate power transmission grid as a prerequisite.
Given the protracted nature of infrastructure development, we must upgrade our power grids today to prepare for a low-carbon future.
Failure to do so may undermine global climate goals.

\section*{Methods}

\subsection*{Power flow optimization models.}
To quantify the reduction in CO$_2$ emissions that can be achieved by the electrification of motor vehicles, we need a mathematical model to describe the operational state of the power grid before and after the mass integration of EVs. The model should be able to determine the maximum carbon reduction achievable from the EV transition under various technical constraints on the power grid operation and reveal potential factors that may hamper further reduction. The model consists of two stages: the first stage (Model I) determines the 
economic operation
of the power grid without EV integration, whereas the second stage (Model II) captures the impact of EV integration on grid operation and emissions. 

Model I is given by the following multi-period optimal power flow problem:

\begin{equation}\label{firststagemodel_obj}
	\underset{    
	\{p_{{\rm g}j} (t) \}_{j \in \mathcal{G},~ t\in \mathcal{T}}	 
	}{\text{\rm min}} \ \  \  \sum_{t\in \mathcal{T}} \sum_{j\in \mathcal{G}} c_{{\rm g}j}\left( p_{{\rm g}j}(t) \right)\Delta t \hspace{0.8in}
	\end{equation}      

		\begin{align}[left = \text{s.t.} \empheqlbrace\,]    
		&  (1-\tau)  \sum_{j\in \mathcal{G}} p_{{\rm g}j}(t) = \sum_{i\in \mathcal{N}} p_{{\rm d}i}(t),~ \forall t \in \mathcal{T}  \label{con0},\\
		& \Big | \sum_{j\in \mathcal{G}} \pi_{lj}^{\rm g} p_{{\rm g}j}(t) - \sum_{i\in \mathcal{N}} \pi_{li}^{\rm d} p_{{\rm d}i}(t) \Big |   \leq F_l, \  \forall l\in \mathcal{L}, \  t\in \mathcal{T} \label{con1},\\
		& 0 \leq  p_{{\rm g}j}(t) \leq  p_{{\rm g}j}^{\rm max}(t), \forall j\in \mathcal{G}, \ t\in \mathcal{T} \label{conv},\\
		& -r_{{\rm g}j}^{\rm down}  \leq p_{{\rm g}j}(t)-p_{{\rm g}j}(t-1) \leq r_{{\rm g}j}^{\rm up},~ \forall j\in \mathcal{G}, \ t\in \mathcal{T}. \label{con2}
		\end{align}
Here, $\mathcal{T}$ represents the set of time steps of an operational cycle.
In this study, we consider an operational cycle of 24 hours, with each hour as a time step in the model; that is, $\mathcal{T}=\{1,2,\cdots,24\}$ and  $\mathrm{\Delta} t = 1$ h. The notations $\mathcal{N}$, $\mathcal{L}$, and $\mathcal{G}$ represent the set of all nodes, transmission lines, and 
generation units in the power grid, respectively. Meanwhile, $p_{{\rm d}i}(t)$ represents the load demand at node $i$ and time $t$, whereas $p_{{\rm g}j}(t)$ stands for the  power generated by unit $j$ at time $t$. The function $c_{{\rm g}j}(\cdot)$ describes the generation cost of generating unit $j$. 
The parameter $p_{{\rm g}j}^{\rm max}(t)$ represents the maximal power that can be generated by unit $j$ at time $t$. The constant $\tau$ accounts for the transmission losses in the network. For conventional generators,  $p_{{\rm g}j}^{\rm max}(t)$ is constant over time, whereas for solar and wind farms, this parameter depends on the weather conditions and is thus time-varying. The constants $r_{{\rm g}j}^{\rm down}$ and $r_{{\rm g}j}^{\rm up}$ capture the respective maximal ramp down and ramp up speeds of generation unit $j$. Moreover, the parameter $\pi_{li}^{\rm d}$ is the power transfer distribution factor (PTDF) from node $i$ to transmission line $l$ and it describes the incremental change of power flow on line $l$ if a unit of net power injection is added (or the net power demand is reduced by one unit) at node $i$. For more details on the calculation of PTDF based on network data, see refs.~[\citenum{wood2013power,zimmerman2010matpower}]. Similarly, $\pi_{lj}^{\rm g}$ represents the PTDF from generator $j$ to transmission line $l$. Finally, the parameter $F_l$ represents the maximal power that can be delivered by transmission line $l$. The optimization model given by Eqs.~(\ref{firststagemodel_obj})-
(\ref{con2})
seeks an optimal generation strategy $p^*_{{\rm g}j}(t), j \in \mathcal{G}, t\in \mathcal{T}$ that minimizes the total generation cost to meet the load demand at every time step (\ref{con0}) while respecting the transmission capacity constraints (\ref{con1}), the generation capacity constraints (\ref{conv}), and the generation ramp-rate constraints (\ref{con2}). 

Model II accounts for the integration of EV charging and is formulated as the following optimal re-dispatch/charge problem: \vspace{-0.5cm}
    
\begin{equation}\label{secondagemodel_obj}
	\underset{    \{ \Delta p_{{\rm g}j}(t),~ p_{\text{v}k}(t) \}_{j \in \mathcal{G},~ k \in \mathcal{V},~t\in \mathcal{T}}
	}{\text{\rm min}} \ \  \ \sum_{t\in \mathcal{T}} \sum_{j\in \mathcal{G}} e_{{\rm g}j}\left( p^*_{{\rm g}j}(t)+\Delta p_{{\rm g}j}(t) \right)\Delta t 
	\end{equation}    
    
		\begin{align}[left = \text{s.t.} \empheqlbrace\,]    
		&  (1-\tau)  \sum_{j\in \mathcal{G}} \left( p^*_{{\rm g}j}(t)+\Delta p_{{\rm g}j}(t) \right) = \sum_{i\in \mathcal{N}} p_{{\rm d}i}(t) +  \sum_{k\in \mathcal{V}}  p_{\text{v}k}(t), \ \forall t \in \mathcal{T}  \label{con01},\\
		& \Big | \sum_{j\in \mathcal{G}} \pi_{lj}^{\rm g} \left( p^*_{{\rm g}j}(t)+\Delta p_{{\rm g}j}(t) \right) - \sum_{i\in \mathcal{N}} \pi_{li}^{\rm d} p_{{\rm d}i}(t) - \sum_{k \in \mathcal{V}}  \pi_{lk}^{\rm v}    p_{\text{v}k}(t) \Big |   \leq F_l, \  \forall l\in \mathcal{L}, \  t\in \mathcal{T} \label{con11},\\
		&  p^*_{{\rm g}j}(t)+\Delta p_{{\rm g}j}(t)  \leq  p_{{\rm g}j}^{\rm max}(t), \ \forall j\in \mathcal{G}, \ t\in \mathcal{T} \label{conv1},\\
		& -r_{{\rm g}j}^{\rm down}  \leq  p^*_{{\rm g}j}(t)+\Delta p_{{\rm g}j}(t) - p^*_{{\rm g}j}(t-1)-\Delta p_{{\rm g}j}(t-1)  \leq r_{{\rm g}j}^{\rm up}, \ \forall j\in \mathcal{G}, \ t\in \mathcal{T} \label{con21},\\
		& \Delta p_{{\rm g}j}(t) \geq 0 , \ \forall j\in \mathcal{G}, t\in \mathcal{T} \label{con51},\\
		&\sum_{k\in \mathcal{V}_c} \sum_{t\in \mathcal{T}} p_{\text{v}k}(t) \Delta t = E_{c}, \ \forall c \in \mathcal{C}  \label{con31},\\
		& p_{\text{v}k}(t) \geq 0 , \ \forall k\in \mathcal{V}, t\in \mathcal{T}, \label{con41}
		\end{align}

\noindent
where we recall that $p^*_{{\rm g}j}(t)$ is the solution of Model I for the base power flow. The new terms include
$\Delta p_{{\rm g}j}(t)$, which is the re-dispatch of generator $j$ in response to EV integration, and  the decision variable $p_{\text{v}k}(t)$, which is the power consumption of the EV charging station $k$ at time $t$. 
In the objective function, $e_{{\rm g}j}(\cdot)$ is the $\text{CO}_2$ emissions function of generation unit $j$. The sets $\mathcal{C}$, $\mathcal{V}_c$, and $\mathcal{V}$ represent the counties, the EV charging stations in county $c$, and the set of all charging stations across the country, respectively.
The parameter
 $\pi_{lk}^{\rm v}$ represents the PTDF from charging station  $k$ to transmission line $l$, whereas  $E_{c}$ denotes the energy requirement of EVs in county $c$ during the entire operational cycle.
Therefore, 
given a solution to Model I, Model II seeks to identify an optimal generation adjustment  $\mathrm{\Delta} p_{{\rm g}j}^*(t), \ j\in \mathcal{G},\ t \in \mathcal{T}$ and EV charging strategy $p_{\text{v}k}^*(t), \ k\in \mathcal{V},\ t \in \mathcal{T}$ that minimize the aggregate $\text{CO}_2$ emissions of the generation units while meeting the base load and EV energy demand. This solution also respects various technical constraints imposed by the transmission network and generation units.

Once we obtain the solutions of Models I and II, the carbon emissions from EV integration can be calculated as follows: 
\vspace{-0.5cm}

\begin{equation}
    e_{\rm EV} = \sum_{t\in \mathcal{T}} \sum_{j\in \mathcal{G}} e_{{\rm g}j}\left( p^*_{{\rm g}j}(t)+\Delta p_{{\rm g}j}^*(t) \right)\Delta t - \sum_{t\in \mathcal{T}} \sum_{j\in \mathcal{G}} e_{{\rm g}j}\left( p^*_{{\rm g}j}(t)\right)\Delta t.
\end{equation}
The total carbon emissions from all vehicles are the sum of those from EVs and the tailpipe emissions from ICVs, $
    e_{\rm V} =  e_{\rm EV} +  e_{\rm ICV}.
    $

\subsection*{Transmission line upgrade model.}  
Model III recognizes that network capacity is a key limiting factor for carbon reduction from vehicle electrification and thus seeks to identify the best strategy to upgrade the transmission lines. 
Specifically, the model is formulated as an optimization problem in which the capacity upgrade required to reduce  CO$_2$ emissions from EVs is minimized:
\vspace{-0.5cm}

\begin{equation}\label{upgrademodel_obj}
	\underset{   \{\Delta F_l \}_{l\in \mathcal{L}},~ \{ \Delta p_{{\rm g}j}(t),~ p_{\text{v}k}(t)\}_{j \in \mathcal{G},~ k \in \mathcal{V},~  t\in \mathcal{T}}
	}{\text{\rm min}} \ \  \ \sum_{l\in \mathcal{L}}  \Delta F_l \cdot m_l \hspace{2in}
	\end{equation}     
    
		\begin{align}[left = \text{s.t.} \empheqlbrace\,]    
		& (1-\tau) \sum_{t\in \mathcal{T}} \sum_{j\in \mathcal{G}} e_{{\rm g}j}\left( p^*_{{\rm g}j}(t)+\Delta p_{{\rm g}j}(t) \right)\Delta t - \sum_{t\in \mathcal{T}} \sum_{j\in \mathcal{G}} e_{{\rm g}j}\left( p^*_{{\rm g}j}(t)\right)\Delta t \leq e_{\rm EV}^{\rm max}  \label{concarbon},\\
		&    \sum_{j\in \mathcal{G}} \left( p^*_{{\rm g}j}(t)+\Delta p_{{\rm g}j}(t) \right) = \sum_{i\in \mathcal{N}} p_{{\rm d}i}(t) +  \sum_{k\in \mathcal{V}}  p_{\text{v}k}(t), \forall t \in \mathcal{T}  \label{con02},\\
		& \Big | \sum_{j\in \mathcal{G}} \pi_{lj}^{\rm g} \left( p^*_{{\rm g}j}(t)+\Delta p_{{\rm g}j}(t) \right) - \sum_{i\in \mathcal{N}} \pi_{li}^{\rm d} p_{{\rm d}i}(t) - \sum_{k \in \mathcal{V}} \pi_{lk}^{\rm v}    p_{\text{v}k}(t) \Big |   \leq F_l + \Delta F_l, \nonumber \\ & \hspace{11cm} ~ \forall l\in \mathcal{L}, \  t\in \mathcal{T},   \label{con12}\\
		&   p^*_{{\rm g}j}(t)+\Delta p_{{\rm g}j}(t)  \leq  p_{{\rm g}j}^{\rm max}(t), \forall j\in \mathcal{G}, \ t\in \mathcal{T} \label{conv2},\\
		& -r_{{\rm g}j}^{\rm down}  \leq  p^*_{{\rm g}j}(t)+\Delta p_{{\rm g}j}(t) - p^*_{{\rm g}j}(t-1)-\Delta p_{{\rm g}j}(t-1)  \leq r_{{\rm g}j}^{\rm up}, \forall j\in \mathcal{G}, \ t\in \mathcal{T} \label{con22},\\
		& \Delta p_{{\rm g}j}(t) \geq 0 , \ \forall j\in \mathcal{G}, t\in \mathcal{T} \label{con52},\\
		& \sum_{k\in \mathcal{V}_c} \sum_{t\in \mathcal{T}} p_{\text{v}k}(t) \Delta t = E_{c},\ \forall c \in \mathcal{C}   \label{con32},\\
		& p_{\text{v}k}(t) \geq 0 , \ \forall k\in \mathcal{V}, t\in \mathcal{T}. \label{con42}
		\end{align}
 
 \noindent
The parameter $m_l$ is the length 
 and   $\mathrm{\Delta}  F_l$ is the incremental capacity increase of transmission line $l$ through the upgrading strategy. The goal of the optimization in equation~(\ref{upgrademodel_obj}) is to minimize the total line capacity increase in MW$\cdot$mile
 needed to reduce the carbon emission from EVs to no more than $e_{\rm EV}^{\rm max}$, as specified by the constraint in equation~(\ref{concarbon}). 
 The other equations represent the same technical constraints that appear in Model II. All three models are linear programs that can be solved by the Gurobi optimization solver.

  All three optimization models adopt the DC power flow approximation to represent the power grid physics, yielding computationally tractable linear programs. As the industrial standard for power system economic operation \cite{zhu2015optimization}, market clearing \cite{biggar2014economics}, and long-term planning  \cite{choi2021probabilistic}, this formulation focuses on active power flow while intentionally neglecting reactive power effects, voltage disparities, and other nonlinearities inherent to AC power flow models. This simplification remains appropriate for our economic and emissions analysis, where active power distribution is the primary concern.

\subsection*{Data sources and assumptions.} 
To instantiate the above models for the EV transition in the U.S., we curated data from various sources on the transmission network, power plants, renewable generation, and vehicle fuel consumption. The U.S. power grid data are reported in the FERC Form 715 \cite{FERC}. The FERC data used in this work contain information on the power network (topology and impedances), generation units (network positions and capacities), and load nodes (network positions and power demands) for the summer peak of 2018. Based on this information, we  calculated the PTDFs ($\pi_{li}^{\rm d}$, $\pi_{lj}^{\rm g}$, $\pi_{lk}^{\rm v}$) 
from each load, generator, and charging station (once we have specified the network locations of charging stations) to each transmission line for all three interconnections. The FERC data only contain peak power demands, but we needed the hourly load curve for the operational cycle of a day under study. The hourly load demand data were obtained from the EIA-930 dataset \cite{EIA930}, which includes the historical hourly aggregate load demand for each region in a $13$-region partition of the system. Three of these regions belong to the Western interconnection (i.e., Northwest, California, and Southwest), one represents Texas, and the others form the Eastern interconnection (Central, Midwest, Mid-Atlantic, Tennessee, New England, New York, Carolinas, Southeast, and Florida). Using the 2018 EIA-930 data, we took the average hourly load curve over all days of a month as the representative hourly load curve for a typical day in that region during that month. 
In this way, we obtained $13\times 12$ load curves, each with $24$ segments. We then normalized each of these load curves by the corresponding peak value over the whole year in the respective region. The normalized load curves were then used to scale the power demand of each load in the FERC data according to the region where the load is located, based on data from the Energy Visuals Transmission Atlas (EVTA) \cite{Atlas}.
This resulted in $12$ hourly load curves $p_{{\rm d}i}(t),\ t=1,2,\cdots,24,$ for each load,  representing typical daily load profiles across the $12$ months of 2018. After obtaining the complete load profile for 2018, we further scaled the load curves according to the annual growth rate (0.55\%) reported in the EIA Annual Energy Outlook 2019 \cite{AEO19}, which enabled us to create a projected load profile for 2025. 

For the generation units, information on their primary energy sources, generation costs, emission rates, and geographic locations was obtained from a combination of data providers. The relevant data on primary source and cost function $c_{{\rm g}j}(\cdot)$ of each generation unit were acquired from the FirstRate Generator Cost Database \cite{FirstRate}. Data on the $\text{CO}_2$ emission rate $e_{{\rm g}j}(\cdot)$ were compiled for each generation unit
using the EPA Emissions \& Generation Resource Integrated Database (eGRID) \cite{egrid}. 
Geographic information was obtained from EVTA \cite{Atlas}. 
For conventional generators, the maximum power generation $p_{{\rm g}i}^{\rm max}(t)$ is equal to the generator capacity and is constant over time. 
However, for wind and solar farms, the maximum power generation at any instant depends on the wind and radiation conditions, respectively, and is upper-bounded by their installed capacity. To complete our optimization models, we needed hourly wind speed and radiation strength data for each wind and solar farm. The hourly wind speed data for the geographic location of each wind farm are available from the NREL Wind Integration National Dataset through the Wind Toolkit Application Programming Interface (API) \cite{windtool}. By assuming that the wind farms follow a typical generation characteristic curve, namely a cubic function with a cut-in speed of $3$ms$^{-1}$ and a cut-off speed of $15$ms$^{-1}$, we converted the hourly wind speed into the corresponding hourly power generation per unit capacity. For solar farms, the hourly power generation per unit capacity at each geographic location is available from the NREL National Solar Radiation Database through the PVWatts API \cite{solardata}. 
The hourly power generation profile $p_{{\rm g}i}(t)$ for each wind and solar farm was then obtained by multiplying the unit-capacity power generation by the capacity of the respective plant. Our analysis also used that the transmission loss rate in the U.S. power grid is $\tau = 0.05911$ according to the most recent data from IEA Statistics \cite{worldbank}.

To further model the transition from ICVs to EVs, we obtained the state-wise motor fuel consumption data from the  FHWA Highway Statistics 2019 \cite{FHA} (which concerns data for the year 2018). 
As an estimation of the county-level data, we then distributed the motor fuel consumption of each state across all counties in proportion to their populations. This proportional distribution approach introduces an estimation error typically small, as indicated by comparison with county-level vehicle registration data available for the 40 most populous counties in California. When considering the transition to EVs, we assumed that the average fuel efficiency of ICVs is 26.0 MPGe (the average real-world fuel efficiency for Model Year 2017 reported in the EPA 2019 Automotive Trends Report \cite{EPAtrend}) and the average efficiency for EVs is $98.2$ MPGe (the average real-world efficiency of Tesla for Model Year 2017 reported in the EPA 2019 Automotive Trends Report). 
In this way, for any percentage of EV transition (we considered $0\%$, $20\%$, $40\%$, $60\%$, $80\%$, $100\%$), we could calculate the daily electricity demand from EVs in each county and hence identify the parameter $E_c$ in Models II and III.
As end-use loads, EV charging stations typically connect at distribution-level voltages. 
Given the absence of detailed distribution network data in the FERC dataset, we model EV charging demand by assigning each county's aggregate charging load to network nodes below 200 kV (subtransmission- and distribution-level voltages) within the same county.  This approach explicitly defines the network locations of charging stations (denoted as $\mathcal{V}_c$)  for each county $c\in \mathcal{C}$.
To project the fuel consumption of 2018 to 2025, we assumed that the motor fuel consumption under 0\% EV integration would grow at the same rate as the U.S. population (0.58\% annually according to U.S. Census Bureau \cite{census}). Based on the data and assumptions described above, Models I,  II, and  III are all well-defined and can be solved numerically.

\section*{Data availability}

This study is based on datasets that are publicly available from the following sources referenced in Methods: the U.S. Energy Information Administration (EIA), U.S. Environmental Protection Agency (EPA), National Renewable Energy Laboratory (NREL), Federal Highway Administration (FHWA), and U.S. Census Bureau. Additionally, the study incorporates data from the Federal Energy Regulatory Commission (FERC) and Energy Visuals, obtained under a nondisclosure agreement following the procedure described at \url{www.ferc.gov/legal/ceii-foia/ceii.asp}.
Data required to generate figures in this paper are provided in our Code Ocean capsule (\url{codeocean.com/capsule/4153860}), where non-publicly available data are shared in processed form, in accordance with nondisclosure requirements. Lastly, we confirm that the views and opinions expressed herein are those of the authors and do not necessarily reflect those of the United States government or any of its agencies.

\section*{Code availability}

The custom code used for implementing the optimization models and generating the figures is available from the Code Ocean capsule (\url{codeocean.com/capsule/4153860}).

\section*{Acknowledgements}

This research was supported by Leslie and Mac McQuown through the Center for Engineering Sustainability and Resilience and by a Resnick Award from the Institute for Sustainability and Energy at Northwestern (now the Paula M.\ Trienens Institute for Sustainability and Energy). C.D.\ also acknowledges support from the National Natural Science Foundation of China under Grant No.\ GQQNKP001.

\section*{Authors' contributions Statement}

C.D.\ and A.E.M.\ designed the research. C.D.\ completed the simulations. C.D.\ and A.E.M.\ analyzed the data and wrote the manuscript.

\section*{Competing interests Statement}

The authors declare no competing interests.



\end{document}